\documentclass[10pt]{elsart}

\usepackage[T1]{fontenc}

\usepackage{graphicx}

\begin{document}

\runauthor{Rocha Filho}
\begin{frontmatter}
\title{Solving the Vlasov equation for one-dimensional models with long range interactions on a GPU}
\author[Rocha]{Tarc\'\i sio M.\ Rocha Filho\thanksref{X}}

\address[Rocha]{Instituto de F\'\i{}sica and International Center for Condensed Matter Physics\\ Universidade de
Bras\'\i{}lia, CP: 04455, 70919-970 - Bras\'\i{}lia, Brazil}
\thanks[X]{e-mail: marciano@fis.unb.br}
\begin{abstract}
We present a GPU parallel implementation of the numeric integration of the Vlasov equation in one spatial dimension
based on a second order time-split algorithm with a local modified cubic-spline interpolation. We apply our approach
to three different systems with long-range interactions: the Hamiltonian Mean Field, Ring and the self-gravitating
sheet models. Speedups and accuracy for each model and different grid resolutions are presented.
\end{abstract}
\end{frontmatter}

\section{Introduction}

Systems with long-range interactions are particularly important in physics, Coulomb forces being probably the most prominent example.
Albeit their relevance, many of its properties are still not well understood.
The long-range nature of the interaction leads to some interesting phenomena not observed for short-range interactions, such as
the existence of quasi-stationary non-Gaussian states with diverging life-times with the number of particles,
negative microcanonical heat capacity, inequivalence of ensembles and non-ergodicity~\cite{prrev,proc1,proc2,proc3,nos,eplnos,benetti}.
Examples of systems with long range forces include
self-gravitating systems~\cite{1b}, non-neutral plasmas~\cite{1c,1d} and models as the ring model~\cite{ring1,ring2}
Hamiltonian Mean Field (HMF)~\cite{hmforig}, one-dimensional gravity (infinite uniform density sheets)~\cite{sheet1,sheet2,sheet3}, Free Electron Laser~\cite{felmod}
and plasma single wave models~\cite{tennyson}, among others.
For out of equilibrium situations, many of these studies rely on molecular dynamics simulations, i.~e.\ solving numerically the Hamiltonian
equations of motion for the $N$-particle system. It is also a well known fact that, under suitable conditions, the statistical description of
the dynamics of long range interacting systems is equivalent to the Vlasov equation~\cite{prrev,braun}. The numerical solution of the Vlasov equation
was applied to the HMF model in Ref.~\cite{antoniazzi} and more
recently to characterize non-equilibrium phase-transitions in the same model~\cite{debuyl1,debuyl2}, although the phase diagram is still open
to a closer scrutiny~\cite{pakter}.

One-dimensional models are important for a better understanding of many properties of long-range interacting systems. Therefore a fast numeric
implementation of the solution of the Vlasov equation is of uttermost value in their investigation.
Numerical solutions of the Vlasov equation are obtained either by Particle In Cell (PIC) methods~\cite{buchner}, where the distribution function
is represented by a collection of macro-particles under the dynamics of the self-consistent mean-field force,
or Eulerian methods where the distribution is represented as the density of a non-compressible fluid on a numerical grid~\cite{pohn}.
For higher dimensional models, PIC methods are more effective in computational effort, even though its applicability
is limited by inherent statistical noise and a poor description of the tails of the distribution.
On the other hand, Eulerian methods are  limited at higher dimensions by the number of grid points required
to accurately represent the distribution function (see~\cite{arber,filbet,shoucri} for a comparison of different Eulerian codes).
With the rapid increase in computational power and the use of parallel machines Eulerian codes have been implemented
up to two spatial dimensions~\cite{crouseilles1,crouseilles2,daldorff,bengt}.

We present in this paper an implementation in the CUDA framework~\cite{cuda} of a semi-Lagrangian solution method for the
Vlasov equation in one dimensional systems with long range interactions, and applications to the ring, HMF and self-gravitating sheet models.
In this approach the distribution function is represented on a
numerical grid and a times-split algorithm is used to evolve the function by computing the characteristic curves and equating the value
of the solution to its value at the foot of the characteristic~\cite{debuyl2,cheng,sonnendrucker}. This last step requires an
interpolation scheme, and a common choice is to use a cubic spline, which is global on the grid
due to the requirement to compute second order derivatives of the distribution
function at grid points~\cite{numrec}. As a consequence its parallel implementation is of limited efficiency.
An alternative in Ref.~\cite{latu}
is to use a local spline on patches (tiles) in the grid, with the continuity of first derivatives at the borders of each patch.
The values of the distribution at the grid points on each patch is stored in shared memory, and all steps are then performed on a
patch-by-patch basis. This approach requires communication between processors handling different patches, which can be reduced
by suitably restricting the time step~\cite{sonnendrucker}.
Here we implement a different approach where the interpolation relies on the same form of cubic spline
with second order derivatives computed from an eighth order finite difference method.

The structure of the paper is as follows: in section~\ref{onedm} we present the one-dimensional models to which our approach is applied,
and section~\ref{paralg} presents and discusses the algorithms implemented in CUDA for the solution of the Vlasov equation.
Section~\ref{resdis} presents the results obtained from the implementation of the algorithm to the one-dimensional models,
and speedups relative to a serial code. We conclude the paper with some concluding remarks in section~\ref{conrem}.

\section{One-dimensional models}
\label{onedm}

Since a detailed direct study of real three-dimensional systems with long-range interactions is a very difficult task, some simplified models have been
introduced in the literature retaining qualitative features of realistic long-range systems (see~\cite{prrev} and references therein).
The Hamiltonian of a one-dimensional model of $N$ identical particles with unit mass can be written as:
\begin{equation}
H=\frac{1}{2}\sum_{i=1}^N p_i^2+\frac{1}{N}\sum_{i<j=1}^N V_{ij},
\label{hamgen}
\end{equation}
where $V_{ij}$ is the potential energy between particles $i$ and $j$ and the factor $N^{-1}$ ensures extensivity of the energy
and corresponds to a change of time units. The three different models considered here correspond to different choices for $V_{ij}$.
The Ring model describes a system of $N$ identical particles on a ring or radius $R$ interacting through their gravitational attraction~\cite{ring1,ring2}.
With a choice of units, the interacting potential is given by
\begin{equation}
V_{ij}=-\frac{1}{\sqrt{2}\sqrt{1-\cos(\theta_i-\theta_j)+\epsilon}},
\label{ringpot}
\end{equation}
where $\theta_i$ is an angle coordinate specifying the position of particle $i$ on the circle, and
$\epsilon$ is a (small) softening parameter used to avoid the divergence of the potential at zero distance. For increasing $\epsilon$
the ring model tends to the HMF model, with potential~\cite{hmforig}:
\begin{equation}
V_{ij}=1-\cos\left(\theta_i-\theta_j \right).
\label{hmfpot}
\end{equation}
The third system considered is the sheet model formed by $N$ identical infinite self-gravitating parallel sheets of constant mass density~\cite{sheet2,sheet3}. 
he gravitational force between two sheets is therefore
constant and they are allowed to cross each other, when the force changes sign. Considering only the motion in the direction $x$ perpendicular to the sheets,
and again with a choice of units, the pair interaction potential is written as:
\begin{equation}
V_{ij}=\left|x_i-x_j \right|.
\label{sheetpot}
\end{equation}

The Vlasov equation for those models is thus:
\begin{equation}
\dot f=\frac{\partial f}{\partial t}+\frac{\partial f}{\partial x}v+\frac{\partial f}{\partial v}F(x,t)=0,
\label{vlasoveq}
\end{equation}
where $f=f(x,v,t)$ is the one particle distribution function, $v$ the velocity of the particle, $x$ the position coordinate
($\theta$ for the HMF and ring models), and the mean-field force $F(x,t)$:
\begin{equation}
F(x,t)=-\frac{\partial}{\partial x}\int  v(x-x')f(x',v',t)\: dx' dv',
\label{meanforce}
\end{equation}
with $v(x-x')$ given in eqs.~(\ref{ringpot}--\ref{sheetpot}). For the HMF model the mean-field force can be written as:
\begin{equation}
F(\theta,t)=-\sin(\theta)M_x+\cos(\theta)M_y,
\label{hmfmeanforce}
\end{equation}
where the components of the ``magnetization'' vector ${\bf M}$ are then
\begin{equation}
M_x=\int  \cos(\theta)f(\theta,v,t)\, dx\:dv,\hspace{5mm}M_y=\int  \sin(\theta)f(\theta,v,t)\: dx\,dv.
\label{mxmydef}
\end{equation}
This property implies that molecular dynamics simulation times for the HMF model with $N$ particles scale as $N$ instead of $N^2$,
and is one the reasons why it is so extensively studied.

\section{Algorithms and CUDA implementation}
\label{paralg}

The semi-Lagrangian scheme used here is described in References~\cite{debuyl2,cheng,sonnendrucker} and can be summarized as follows.
The one-particle distribution function is represented in a numerical grid in the one-particle phase space as
$f(x_i,p_j,t)$ where $x_i$ and $p_j$ are position and velocity coordinates of the points in the grid on a finite domain
$x\in[x_{min},x_{max}]$ and $p\in[p_{min},p_{max}]$.
The distribution function at time $t+\Delta t$ is obtained numerically by evolving the function back in time
and using the invariance of $f$ along the characteristic lines. This backwards evolution is performed using a a time-split method.
The mains steps are:
\begin{enumerate}
\item Backwards time evolution (advection) of $f$ in the spatial direction by a time step $\Delta t/2$ with constant momentum:
\label{st1}
\begin{equation}
f^{(I)}(x,p)=f(x-p\:\Delta t/2,p,t).
\label{step1}
\end{equation}
\item Computation of the mean-field force using $f^{(I)}$:
\label{st2}
\begin{equation}
F^{(I)}(x)=-\int \frac{\partial}{\partial x}v(x-x')f^{(I)}(x',p')\:dx'\:dv'.
\label{step2}
\end{equation}
\item Backwards time evolution in the momentum direction by a full time step $\Delta t$ using $F^{(I)}(x)$:
\label{st3}
\begin{equation}
f^{(II)}(x,p)=f(x,p-F^{(I)}(x)\Delta t,t).
\label{step3}
\end{equation}
\item And as last step repeat~(\ref{st1}):
\label{st4}
\begin{equation}
f(x,p,t+\Delta t)=f^{(II)}(x-p\:\Delta t/2,p,t).
\label{step4}
\end{equation}
\end{enumerate}

The values of the intermediate $f^{(I)}$, $f^{(II)}$ and final distribution functions in steps~(\ref{st1}), (\ref{st3}) and~(\ref{st4}) at
the numerical grid points must be obtained from the known values at the previous step with
a cubic spline as interpolation method. For a point with coordinate $x$, $x_i\leq x\leq x_{i+1}$, the interpolated value for $f(x)$
knowing $f_i=f(x_i)$ and $f_{i+1}=f(x_{i+1})$ is given by~\cite{numrec}:
\begin{equation}
f(x)=\alpha f_i+\beta f_{i+1}+\gamma f^{\prime\prime}_i+\delta f^{\prime\prime}_{i+i},
\label{splinegen}
\end{equation}
where
\begin{eqnarray}
 & & \alpha=\frac{x_{i+1}-x}{x_{i+1}-x_i},\hspace{3mm}\beta=1-\alpha,\nonumber\\
 & & \gamma=\frac{\alpha^3-\alpha}{6}(x_{i+1}-x_i)^2,
\hspace{3mm}\delta=\frac{\beta^3-\beta}{6}(x_{i+1}-x_i)^2,
\label{splinecoeffs}
\end{eqnarray}
and $f^{\prime\prime}_i$ stands for the second derivative of $f$ at $x_i$.
BY requiring that first order derivatives computed from eq.~(\ref{splinegen}) are continuum across the boundaries of neighboring
intervals, we obtain a tridiagonal system of equations for $f_i^{\prime\prime}$:
\begin{equation}
\left(\frac{f_{i+1}^{\prime\prime}}{6}+\frac{f_i^{\prime\prime}}{3}+\frac{f_{i-1}^{\prime\prime}}{6}\right)\Delta x^2=f_{i+1}-2f_i+f_{i-1},
\hspace{3mm}i=0,\ldots n,
\label{tridiagsys}
\end{equation}
with $n$ the number of points in the corresponding direction. Even though useful
in a sequential context, a direct efficient parallel implementation of the solution of system~(\ref{tridiagsys}) is not effective enough.

In order to compute the second order derivatives locally, i.~e.\ involving only a small number of neighbor points, with good accuracy in order not
to spoil the quality of the cubic interpolation an eighth order centered finite difference approximation is used~\cite{fornberg}:
\begin{eqnarray}
f^{\prime\prime}(u_i) & = & -\frac{1}{560}f(u_{i-4})+\frac{8}{315}f(u_{i-3})-\frac{1}{5}f(u_{i-2})+\frac{8}{5}f(u_{i-1})-\frac{205}{72}f(u_i)\nonumber\\
 & & +\frac{8}{5}f(u_{i+1})-\frac{1}{5}f(u_{i+2})+\frac{8}{315}f(u_{i+3})-\frac{1}{560}f(u_{i+4}),
\label{findif}
\end{eqnarray}
where $u$ stands for either the momentum or position variables. Periodic boundary conditions are used both in the spatial and momentum direction.
Unphysical effects are avoided by choosing the size of the domain sufficiently large.

The distribution function is represented on an equally spaced grid  $(x_i,p_j)$, with $n_x\times n_p$ points, by
a one-dimensional array $f(x_i,p_j)\rightarrow f[j\cdot n_x+i]$, $i=0,\ldots,n_x$ and $j=0,\ldots,n_p$.
The second order derivatives are represented similarly.
The initial condition array is loaded in
global GPU memory and all subsequent operations are performed there. Memory bandwidth is an important issue for efficiency is this memory intensive
application, and the algorithm must exploit as much as possible coalesced memory access.
Reading and writing in the one-dimensional array in the $x$ direction tends to be coalesced,
but not on the $p$ direction. This is an issue when computing the spline coefficients
(the second order derivatives) necessary for the interpolation in the spatial advection, but is avoided in the other parts of the advection process.
To overcome this difficulty a global transpose of the $f$ array is performed before computing $f^{\prime\prime}$,
and then another transpose is performed it the array $f^{\prime\prime}$.
As a consequence the same routine is used to compute the derivatives for both spatial and momentum directions.
The transpose of a bidimensional array written in the one-dimensional form can be performed efficiently close to
full bandwidth in CUDA~\cite{ruetsch}. Our algorithm is synthesized as:
\renewcommand{\theenumi}{\roman{enumi}}
\begin{enumerate}
\item Transpose $f$;\label{tt1}
\item Compute $f^{\prime\prime}$;\label{tt2}
\item Transpose $f^{\prime\prime}$;\label{tt3}
\item Perform a spatial advection (step~\ref{st1} above) using $f^{\prime\prime}$ obtained in step~(\ref{tt2}) for the spline interpolation;
\item Compute the mean-field force $F^{(I)}$ (step~\ref{st2} above);\label{tt4}
\item Compute $f^{\prime\prime}$;\label{tt5}
\item Perform a momentum advection (step~\ref{st3}) using $f^{\prime\prime}$ from step~\ref{tt5};
\item Repeat steps~(\ref{tt1}--\ref{tt4}) (step~\ref{st4}).
\end{enumerate}
The computation of the force in step~(\ref{tt4}) is implemented by a discretization of eq.~(\ref{meanforce}).

\section{Results and discussions}
\label{resdis}

The simulations presented here were performed on a GTX 560 Ti GPU with 384 cores, 1GB global memory and 1,64 GHz clock speed,
and a GTX 590 GPU with 512 cores, 1,5 GB global memory and 1.26 GHz\ clock (in fact the GTX 590 has two identical devices but
only one was used for the runs).
The CPU has an i7-2600 processor with 3.4GHz clock and 16GB of RAM.
In this section we present and discuss the results of simulations for the three one-dimensional models presented in section~\ref{onedm}.
All computations performed use double precision.

\subsection{Self-gravitating sheet model}

\begin{figure}[ptb]
\begin{center}
\scalebox{0.57}{{\includegraphics{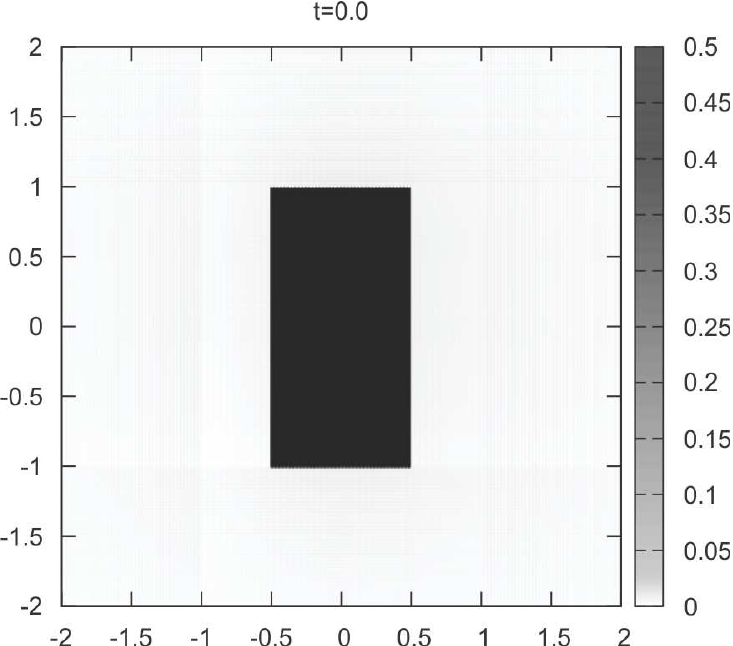}}}
\scalebox{0.57}{{\includegraphics{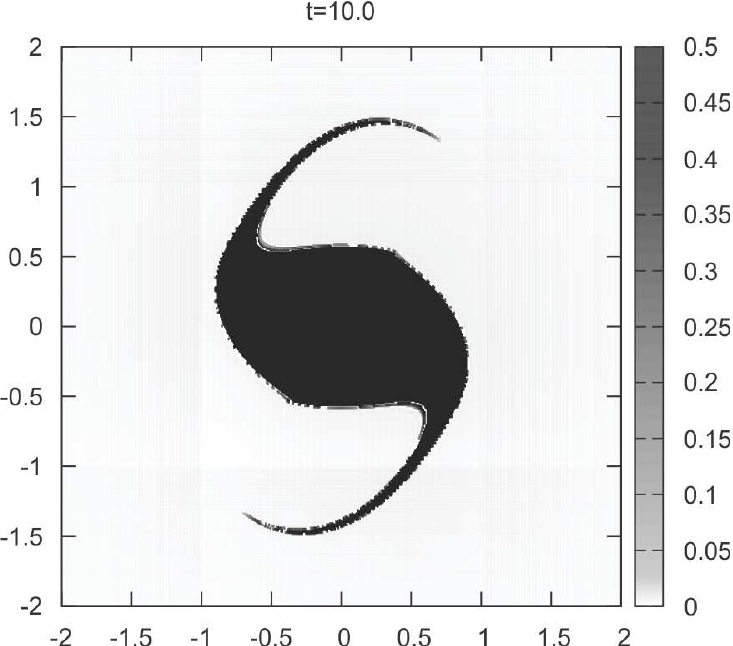}}}
\scalebox{0.57}{{\includegraphics{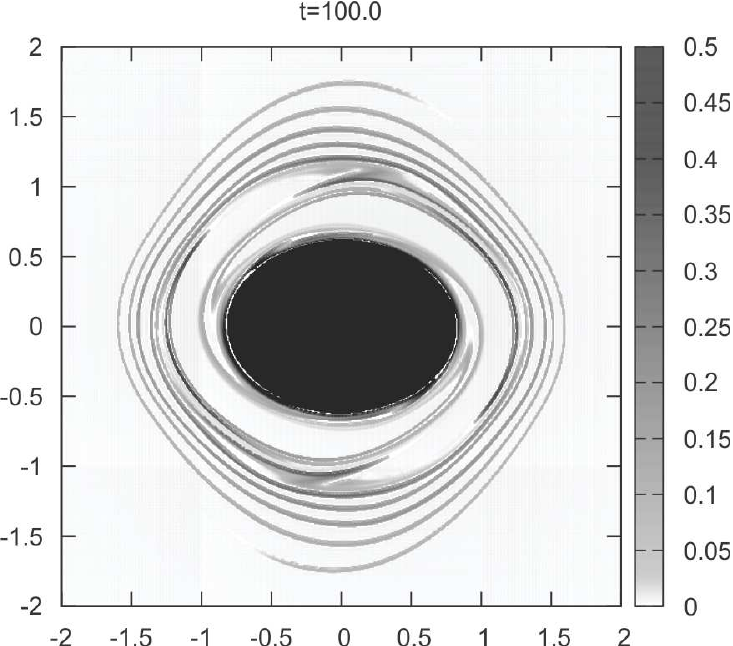}}}
\end{center}
\begin{center}
\scalebox{0.57}{{\includegraphics{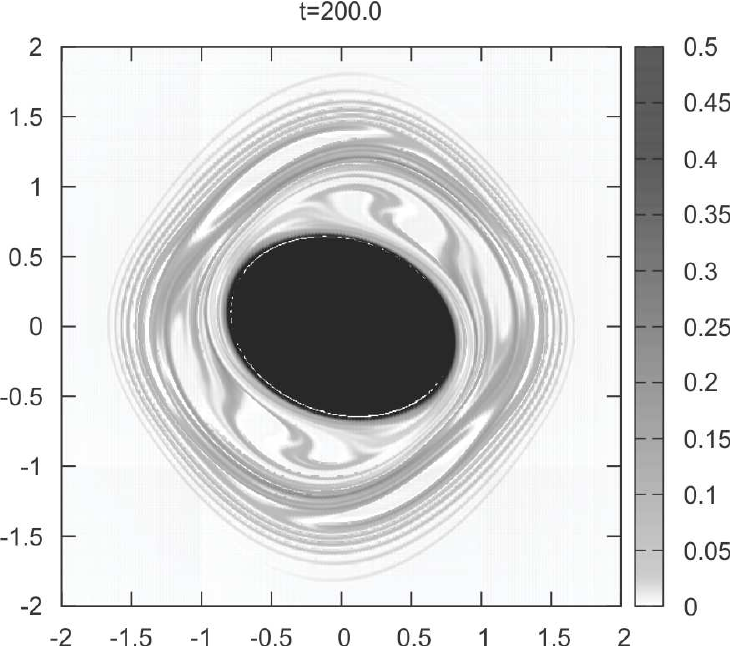}}}
\scalebox{0.57}{{\includegraphics{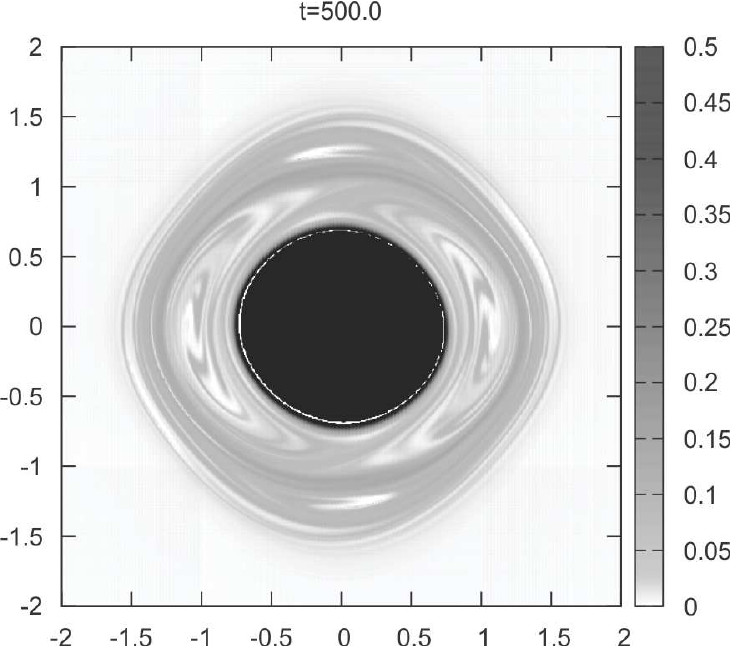}}}
\scalebox{0.57}{{\includegraphics{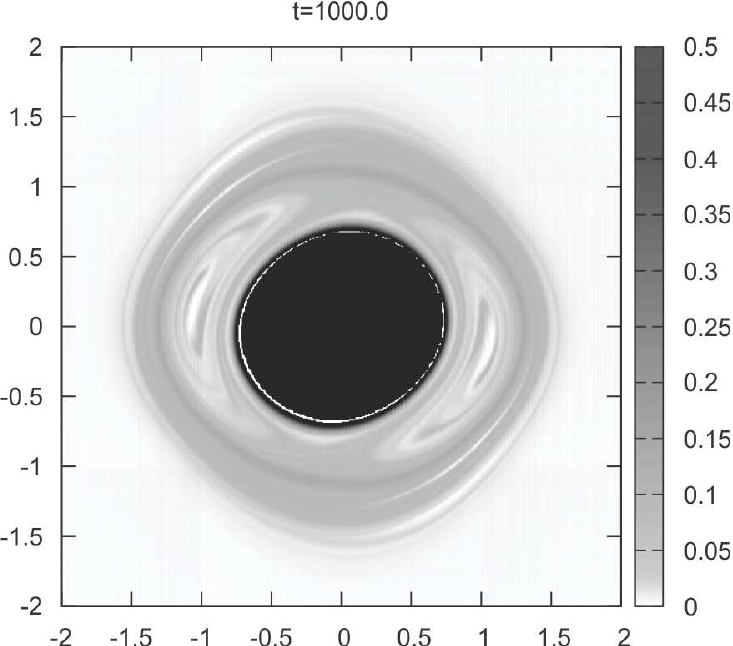}}}
\end{center}
\caption{Snapshots of the distribution function for the self-gravitating sheet model
with $x_{max}=-x_{min}=2.0$, $p_{max}=-p_{min}=2.0$, $n_p=n_x=2048$,
$\Delta t=0.01$, a waterbag initial condition with $x_0=1.0$ and  $p_0=0.5$, and $t=0,\:10,\:100,\:200,\:500,\:1000$.
In each graphic $p$ and $x$ correspond to the horizontal and vertical axis respectively}
\label{fig1}
\end{figure}

As a first test case let us apply our approach to the self-gravitating sheet model with
pair interaction potential given by eq.~(\ref{sheetpot}),
with a (waterbag) constant distribution in an interval as initial condition, i.~e:
\begin{equation}
f(x,p,t=0)=1/4p_0x_0,\hspace{4mm}{\rm if\:\:-x_0\leq x\leq x_0,\:\:and-p_0\leq p\leq p_0}.
\end{equation}
The parameters of the numeric grid are $x_{max}=-x_{min}=2.0$, $p_{max}=-p_{min}=2.0$ and $n_p=n_x=256,512,1024,2048$
(different number of points in each direction can also be used).
The waterbag initial condition is chosen with $x_0=1.0$ and $p_0=0.5$. Two time steps $\Delta=0.1$ and $\Delta t=0.01$ were
considered to assess numerical errors. Figure~\ref{fig1} shows some snapshots of the time evolution of the distribution
function obtained from our code.

Tables~\ref{tab1} and~\ref{tab2} present the relative errors for the energy $\delta e$ and total norm $\delta{Norm}$.
The accuracy of our approach is similar to the a global spline as describe in~\cite{numrec}, for all cases considered.
\begin{table}
\begin{center}
\begin{tabular}{ccccc}
\hline\hline
$n_p=n_x$ & $\delta e^{(\rm FD)}$ & $\delta{Norm}^{(\rm FD)}$ & $\delta e^{(\rm GS)}$ & $\delta {Norm}^{(\rm GS)}$\\
\hline
$256$ & $7\times 10^{-3}$ & $10^{-3}$ & $10^{-2}$ & $10^{-3}$\\
$512$ & $7\times 10^{-4}$ & $10^{-4}$ & $10^{-4}$ & $1.7\times 10^{-5}$\\
$1024$ & $10^{-5}$ & $3\times 10^{-6}$ & $2\times 10^{-6}$ & $9\times 10^{-7}$\\
$2048$ & $4\times10^{-6}$ & $10^{-8}$ & $3\times 10^{-6}$ & $10^{-8}$ \\
\hline
\end{tabular}
\end{center}
\caption{Maximum relative error for the energy ($\delta e$) and norm ($\delta{Norm}$) for the Finite Difference (FD) approximation in eq.~(\ref{findif})
and Global Spline (GS), with $\Delta=0.01$.}
\label{tab1}
\end{table}
\begin{table}
\begin{center}
\begin{tabular}{ccccc}
\hline\hline
$n_p=n_x$ & $\delta e^{(\rm FD)}$ & $\delta{Norm}^{(\rm FD)}$ & $\delta e^{(\rm GS)}$ & $\delta {Norm}^{(\rm GS)}$\\
\hline
$256$ & $5\times10^{-4}$ & $4\times10^{-5}$ & $2\times10^{-4}$ & $4\times10^{-5}$\\
$512$ & $4\times 10^{-4}$ & $2\times 10^{-7}$ & $4\times 10^{-4}$ & $2\times 10^{-7}$\\
$1024$ & $3\times 10^{-4}$ & $3\times 10^{-10}$ & $3\times 10^{-4}$ & $10^{-10}$\\
$2048$ & $3\times 10^{-4}$ & $8\times 10^{-13}$ & $3\times 10^{-4}$ & $10^{-14}$\\
\hline
\end{tabular}
\end{center}
\caption{Same as table~\ref{tab1} with $\Delta t=0.1$.}
\label{tab2}
\end{table}
The speedups obtained by comparing our parallel code to a CPU serial version using the same interpolation method are shown in table~\ref{tab3}.
The speedups grow with $n_p$ and $n_x$ for two main reasons. First not all the latencies of the GPU are covered with a small number of grid points.
And second, with smaller grid spacings the possibility of coalesced access to global memory is significantly increased.
\begin{table}
\begin{center}
\begin{tabular}{ccc}
\hline\hline
$n_p=n_x$ & GTX 570 Ti & GTX 590\\
\hline
\hline
 $256$ & 17 & 20\\
 $512$ & 25 & 31\\
 $1024$ & 38 & 51\\
 $2048$ & 51 & 71\\
\hline
\end{tabular}
\end{center}
\caption{Speedups for the self-gravitating sheet model for different grid resolutions.}
\label{tab3}
\end{table}

\begin{figure}[ptb]
\begin{center}
\scalebox{0.4}{{\includegraphics{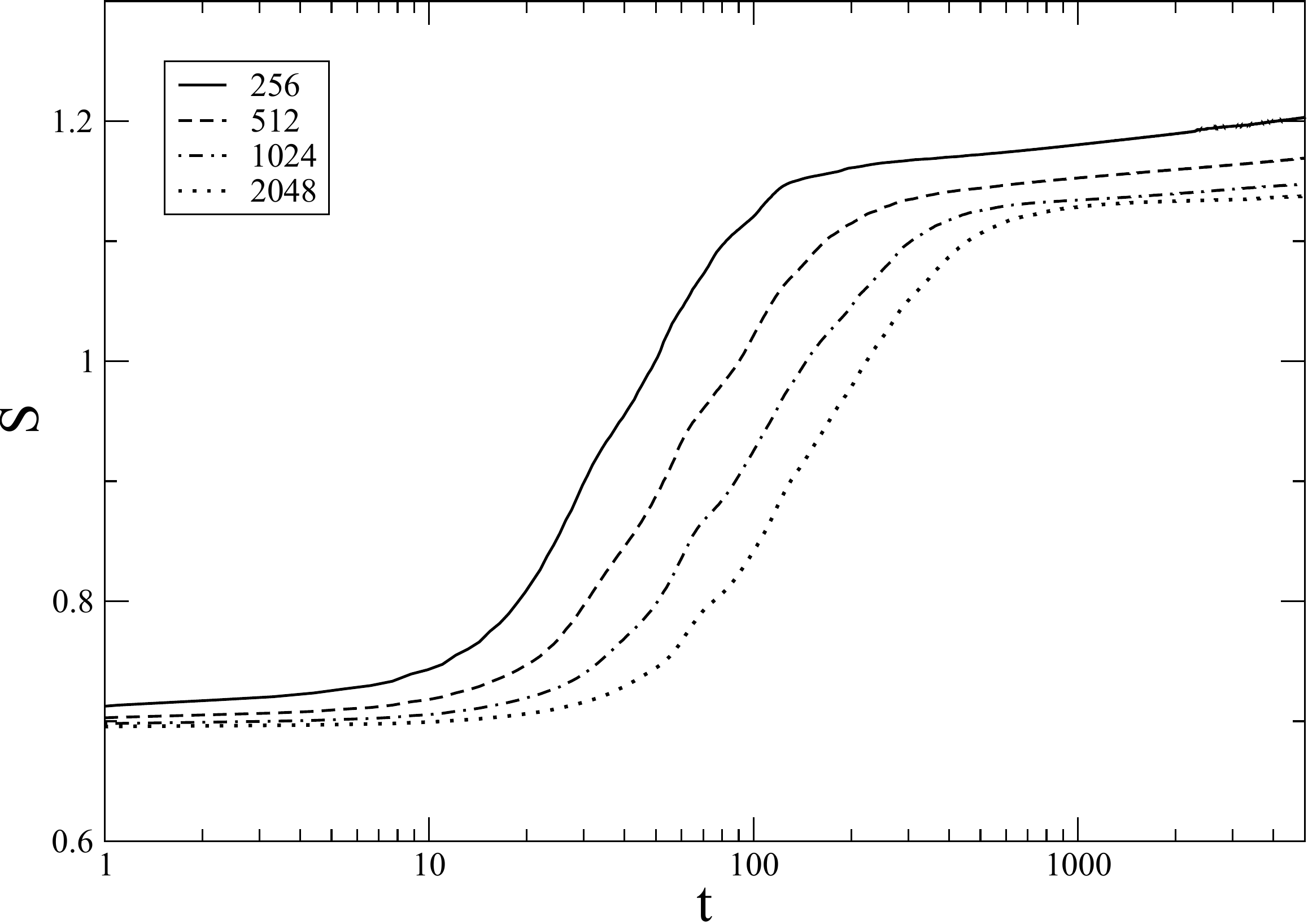}}}
\end{center}
\caption{Entropy for the sheet model for different grid resolutions. The runs are the same as in table~\ref{tab2}.}
\label{sheetent}
\end{figure}

The Vlasov dynamics has an infinite number of invariants, called Casimirs, of the form
\begin{equation}
C[s]=\int s(f(p,x,t))\:dp\:dx.
\label{casimirs}
\end{equation}
This fact can be used to asses how information on the initial condition is lost due to the finite resolution of the numerical grid
and other sources of errors.
For this purpose we consider the entropy of the distribution $f$ given by $s(f)=-f\log f$ in eq.~(\ref{casimirs}).
Figure~\ref{sheetent} shows the time dependence of $S$ for different grid resolutions.
Information loss starts to be significant when filamentation is of the order of the grid spacing.
These non-Vlasov effects are inherent to Eulerian solvers and must be considered
with due care in the numerical solutions of the Vlasov equation~\cite{antoniazzi,galeotti,califano,carbone}.
For the highest resolution ($2024\times2024$), there are are two plateaus, one at the initial stage, before the formation of filamentation,
and another at the final stage, when details smaller than the grid resolutions were lost.

\subsubsection{The HMF model}

\begin{figure}[ptb]
\begin{center}
\scalebox{0.8}{{\includegraphics{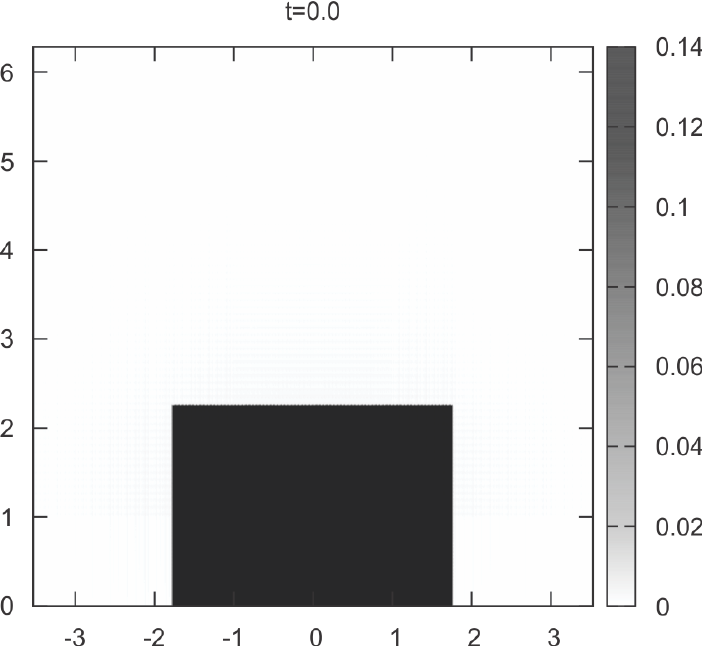}}}
\scalebox{0.8}{{\includegraphics{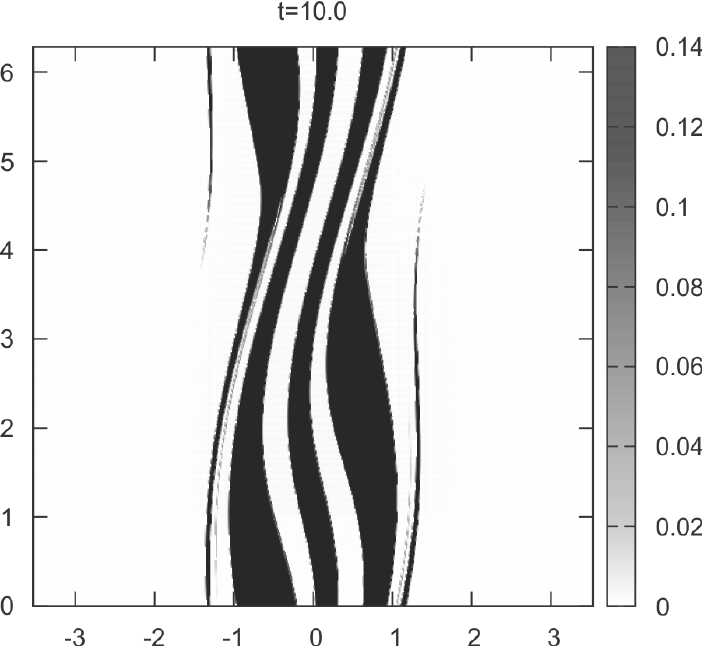}}}
\end{center}
\begin{center}
\scalebox{0.8}{{\includegraphics{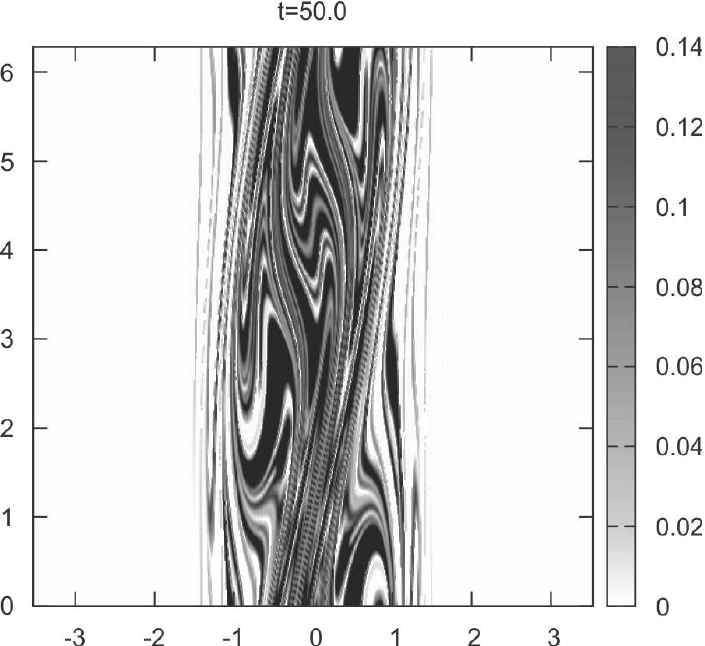}}}
\scalebox{0.8}{{\includegraphics{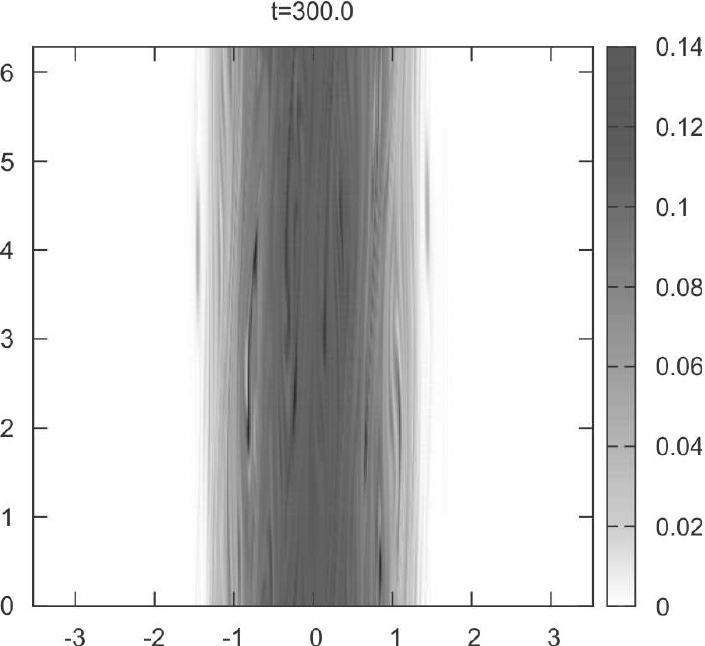}}}
\end{center}
\caption{Snapshots of the distribution function of the HMF model for $t=0,\:10,\:50,\:300$. In each graphic $p$ and $\theta$
correspond to the horizontal and vertical axis respectively.}
\label{hmfsnaps}
\end{figure}

For the HMF model as defined by the pair interaction potential in eq.~(\ref{hmfpot}) Molecular
Dynamics (MD) simulations scale with the number of particles $N$. Therefore it is possible to compare results from MD simulations
with the solutions of the Vlasov equation, which describes the statistical properties of the particle dynamics
in the $N\rightarrow$ limit~\cite{prrev,braun}. As initial condition we consider
a waterbag with total energy per particle $e=0.7$ and average magnetization
$M=\sqrt{M_x^2+M_y^2}=0.8$, with $M_x\equiv\langle\cos(\theta)\rangle$, $M_y\equiv\langle\sin(\theta)\rangle$.
This corresponds to a uniform distribution in the interval $0\leq\theta\leq 2.262$ and $-1.766\leq p\leq 1.766$.
All integrations were performed with $\Delta t=0.1$, a spatial grid $0\leq\theta<2\pi$ and momentum grid
$-p_{max}<p<p_{max}$ with $p_{max}=3.531$ and $n_p=n_\theta=256,512,1024,2048$.
Snapshots of the time evolution of the distribution are shown in Fig.~\ref{hmfsnaps} with a strong filamentation already present at $t=50$.

\begin{figure}[ptb]
\begin{center}
\scalebox{0.27}{{\includegraphics{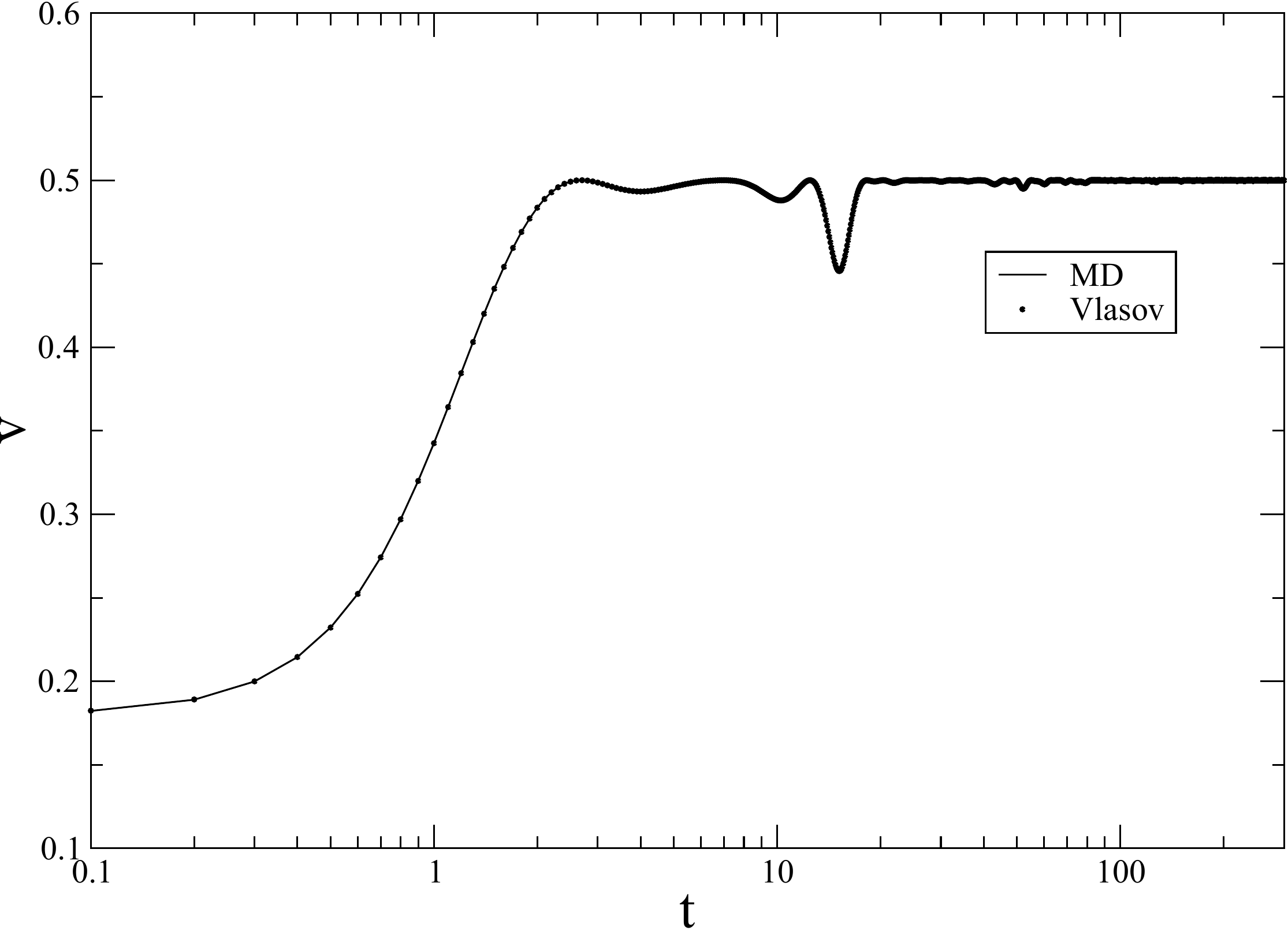}}}
\scalebox{0.27}{{\includegraphics{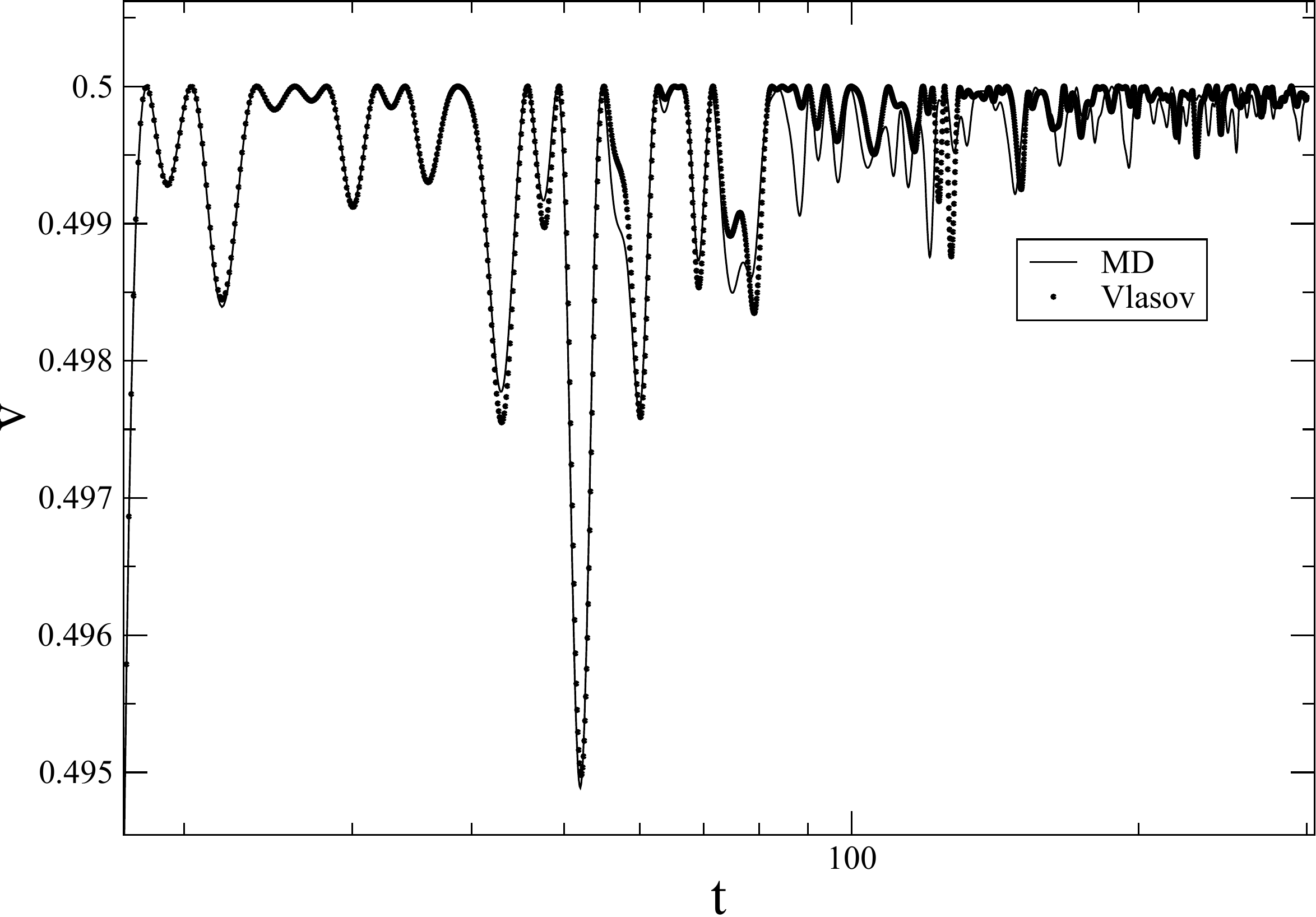}}}
\end{center}
\caption{Mono-Log graphic of the potential energy for the HMF model obtained from Molecular Dynamics simulation (dots)
with $N=20,000,000$ particles
and from the numerical solution of the Vlasov equation with $n_p=n_\theta=2048$ (continuous line). The right panel
is a zoom over a region of the left panel.}
\label{fighmfpot}
\end{figure}

\begin{figure}[ptb]
\begin{center}
\scalebox{0.4}{{\includegraphics{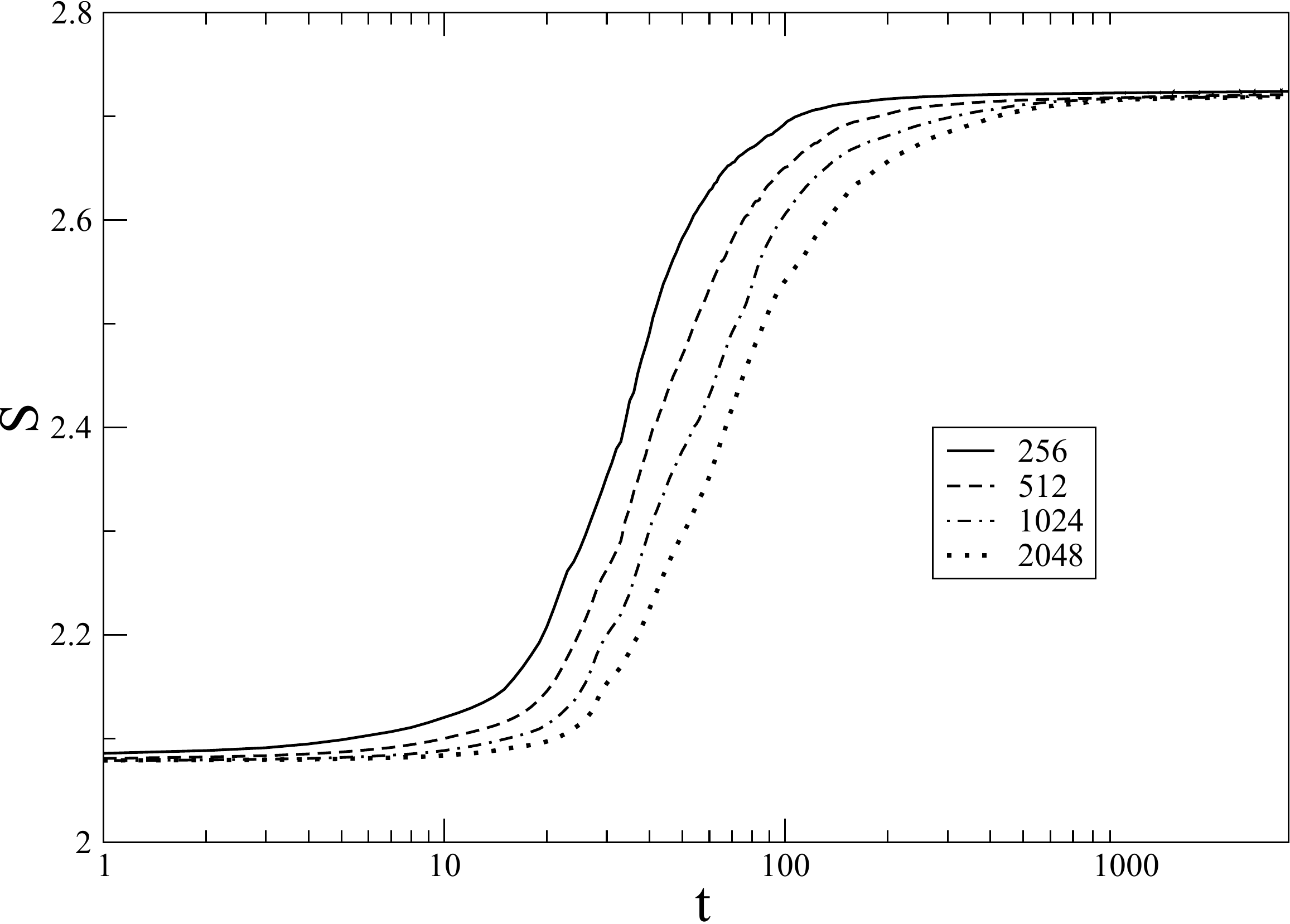}}}
\end{center}
\caption{Entropy for the HMF model for different grid resolutions.}
\label{hmfent}
\end{figure}

Figure~\ref{fighmfpot} shows the graphic of the potential energy obtained from a MD simulation with $N=20,000,000$ particle using
a sympletic integrator with time step $\Delta t=0.1$~\cite{yoshida}, and the same curve obtained from our code with a numerical grid
with $n_p=n_\theta=2048$ points. Both simulations are in very good agreement up to roughly $t\approx 70.0$, after which the details
of small fluctuations differ. This is due to the finite number of particles in the MD simulation and the strong formation
of filaments by the natural evolution of the distribution function down to scales of the size of the grid spacing.
Nevertheless the asymptotic behavior is the same in both cases.
The entropy for different grid resolutions is shown in Fig.~\ref{hmfent}. Analogously to the sheet model, it has two plateaus, one before indentations
scale reaches grid resolution, and a final plateau after the distribution is coarse grained.
The speedups for the HMF model are shown in table~\ref{tab4}, and are somewhat smaller than those for the sheet model. This comes
from the fact that the serial code is well optimized by using eq.~(\ref{hmfmeanforce}). Also the rate of successful coalesced
memory access depends on the dynamics of the model, i.~e.\ how far each grid point is moved by the advection.
Figure~\ref{hmfent} shows the entropy for the HMF model for different number of grid points $n_p=n_\theta$.
The behavior is qualitatively the same as the previous case.

\begin{table}
\begin{center}
\begin{tabular}{ccc}
\hline\hline
$n_p=n_x$ & GTX 570 Ti & GTX 590\\
\hline
\hline
 $256$ & 22 & 24 \\
 $512$ & 29 & 35 \\
 $1024$ & 35 & 37 \\
 $2048$ & 39 & 53 \\
\hline
\end{tabular}
\end{center}
\caption{Speedups for the HMF model for different grid resolutions.}
\label{tab4}
\end{table}

\subsubsection{The Ring model}

Computation of the mean-force field~(\ref{meanforce}) for the pair interaction potential given in eq.~(\ref{ringpot}) involves
the computation of $\sin(\theta-\theta')=\sin\theta\:\cos\theta'-\cos\theta\sin\theta'$, and can therefore be optimized by simply
computing and storing two arrays with the values of $\sin\theta$ and $\cos\theta$ at the spatial grid points at the beginning of
the simulation. The CUDA function {\it rsqrt} for the inverse of the square-root in eq.~(\ref{meanforce}) is exploited as
it is cheaper than to computed both a square root and its inverse. The remaining steps are as described above.
The speedups obtained are presented in table~\ref{tab5}.
The behavior with the number of grid points is similar to the two previous models.
\begin{table}
\begin{center}
\begin{tabular}{ccc}
\hline\hline
$n_p=n_x$ & GTX 570 Ti & GTX 590\\
\hline
\hline
 $256$ & 17 & 19 \\
 $512$ & 29 & 33 \\
 $1024$ & 46 & 55 \\
 $2048$ & 57 & 73 \\
\hline
\end{tabular}
\end{center}
\caption{Speedups for the Ring model model.}
\label{tab5}
\end{table}

\section{Concluding remarks}
\label{conrem}

We presented a GPU implementation using CUDA of a parallel numeric solver for the Vlasov equation on a GPU,
based on a time-split scheme with a modified cubic spline interpolation for both the spatial and momentum direction in phase space.
The interpolation relies on a finite-difference scheme to accurately determine the second order derivatives required by the cubic spline interpolation
in such a a way than only a small number of neighboring points is required,
leading to a faster and simpler parallel implementation. Coalesced access to global memory
in the GPU is ensured by performing a transpose of the distribution function and its second derivatives when performing the
advection in the spatial direction. Implementations for three different one-dimensional long-range interacting models
were presented with a discussion of accuracy and speedups of the simulations. The Vlasov dynamics leads to
the formation of indentations in a scale which becomes smaller with time, and due to the finite grid accuracy, information
loss ensues after some time, leading to a coarse-grained distribution, and an increase in entropy.
Before the scale of indentation reaches grid accuracy the entropy is conserved by our approach. The same occurs after
the distribution has been coarse-grained. Higher order time split schemes and a filtering procedure
can also be implemented if required~\cite{arber}. Although there is certainly room for improvements in our algorithm,
the speedups obtained allow to conclude that the present parallel implementation is a useful tool
in the ongoing investigations on open problems for long-range interacting systems.

\section{Acknowledgments}

The author would like to thank CNPq and CAPES (Brazil) for partial financial support.

\end{document}